\documentstyle[amsfonts,epsfig]{article}

\begin{document}
\title{Entanglement in squeezed two-level atom}
\author{Shigeru Furuichi and Mahmoud Abdel-Aty$^{\dagger}$\\
{\small Science University of Tokyo in Yamaguchi,Onoda 
city,Yamaguchi,756-0884,Japan}\\
$^{\dagger}${\small Mathematics Department, Faculty of Science, South Valley 
University, 82524 Sohag, Egypt}}
\date{ }
\maketitle
{\bf Abstract.}  In the previous paper, we adopted the method using quantum 
mutual entropy to measure the degree of entanglement in the time development 
of the Jaynes-Cummings model \cite{FO}. In this paper, we formulate the 
entanglement in the time development of the Jaynes-Cummings model with 
squeezed states, and then show that the entanglement can be controlled by means of
squeezing.
\vspace{3mm}


\vspace{2mm}


\section{Introduction}
\quad \,

Recently, it has been known that a quantum entangled state plays an 
important role in the field of quantum information theory such as quantum 
teleportation and quantum computation. The research on quantifying entangled 
states has been done by several measures\cite{Ben,Ved}.

If we want to quantify entangled states, we should know whether they are 
pure states or mixed states. That is, if the entangled states are pure 
states, then it is well known that it is sufficient to use von Neumann entropy\cite{Neu}
 for the reduced states\cite{Ben,Ved}. Because, for pure states, it has a unique measure. 
 However, for mixed states, it does not have a unique measure. Therefore we need a proper 
measure of entanglement for mixed state. The degree of entanglement for 
mixed states has been studied by some entropic measures such as entanglement 
of formation\cite{Ben} and quantum relative entropy\cite{Ved} and so on. 
V.Vedral {\it et al.} defined the degree of the entangled states $\sigma$ as 
a minimum distance between all disentangled states $\rho \in {\cal D}$ such 
that $E\left( \sigma \right) \equiv {\min_{_{\rho \in {\cal D}}}}D\left( 
\sigma ||\rho \right) $ where $D$ is any measure of distance between
the two states $\sigma $ and $\rho $. For an example, we can choose
quantum relative entropy as $D$. Then,
\begin{equation}
E\left( \sigma \right) ={\min_{\rho \in {\cal D}}}S\left( \sigma |\rho
\right) ,  \label{QRE}
\end{equation}
where $S\left( \sigma |\rho \right) \equiv tr\sigma \left( \log \sigma -\log
\rho \right) $ is quantum relative entropy\cite{U,OP}. Since this measure 
has
to take a minimum over all disentangled states, it is difficult to calculate
analytically for the actual model such as the Jaynes-Cummings model so that 
we use the degree of entanglement due to mutual entropy\cite{O1}, we 
call it DEM in the sequel, defined below. Moreover, there has been no 
fixed-definition of entanglement measure, though some measures have been 
defined other than the above measure defined in (\ref{QRE}).
So we can use the convenient measure which ever we want case-by-case.

Let $\sigma $ be a state in ${\frak S}_{1}\otimes {\frak S}_{2}$ and $\rho
_{k}$ are the marginal states in ${\frak S}_{k}$ ({\it i.e.}, $tr_{j}\sigma
=\rho _{k}\left( k\neq j\right) $ ). Then our degree of entanglement due to 
mutual entropy (DEM) is defined by:
$$I_{\sigma }\left( \rho _{1},\rho _{2}\right) \equiv tr\sigma \left( \log
\sigma -\log \rho _{1}\otimes \rho _{2}\right).$$
Note that the tensor product state $\rho _{1}\otimes \rho _{2}$ is one of 
the disentangled states. This DEM represents the difference between the entangled
states and disentangled states.
This quantity is also applied to classify entanglement in
\cite{BO}. If we treat only entangled pure states, it is sufficient to apply 
 von Neumann entropy for the {\it reduced} states $\rho _{k} \equiv tr_{j}\sigma \in {\frak S}_{k}, (k\neq j) $. 
 Because we suppose that 
$\sigma \in {\frak S}_{1}\otimes {\frak S}_{2}$ are entangled pure states, 
then its von Neumann entropy is equal to $0$ ($S\left( \sigma \right) =0$). 
Moreover, according to the following triangle inequality of Araki and Lieb\cite{AL}:
$$|S\left( \rho _{1}\right) -S\left( \rho _{2}\right) |\leq S\left( \sigma
\right) \leq S\left( \rho _{1}\right) +S\left( \rho _{2}\right)$$
we have $S\left( \rho _{1}\right) =S\left( \rho _{2}\right) $. Thus we have
\begin{eqnarray}
I_{\sigma }\left( \rho _{1},\rho _{2}\right) &=&tr\sigma \left( \log \sigma
-\log \rho _{1}\otimes \rho _{2}\right)  \nonumber \\
&=&S\left( \rho _{1}\right) +S\left( \rho _{2}\right) -S\left( \sigma \right)
\label{DEM} \\
&=&2S\left( \rho _{1}\right)  \nonumber
\end{eqnarray}
Therefore, for entangled pure states, the DEM becomes twice of the entropy 
of the induced marginal states. That is, if we want to know the degree of the 
entangled pure states, it is sufficient to use the {\it reduced} von Neumann entropy.
However, in general the reduced von Neumann entropies for entangled mixed states are not always unique,
 namely, they depend on the way to take the partial trace, 
 therefore we need the unique measure for the entangled mixed states.
Thus in this paper we apply the DEM not the reduced von Neumann entropy 
in order to measure the degree of entanglement for the entangled mixed states, 
because the DEM can measure the degree of entanglement directly without taking partial trace.
From (\ref{DEM}), we also find that the entanglement degree in pure states 
is bigger than that in mixed states.  In this short paper, we will formulate 
the entanglement degree in the Jaynes-Cummings model with squeezed states 
using DEM and then, try to control it by means of the squeezing parameter 
$r$.

\section{Atomic system}
\quad \,

The quantum electrodynamical interaction of a single two-level atom with a 
single mode of an electromagnetic field is described by the well known 
Jaynes-Cummings model(JCM)\cite{JC}.
The JCM is the simplest nontrivial model of two interacting fully quantum 
systems and has an exact solution. It also brings us some interesting 
phenomena such as collapses and revivals.
It has been investigated in detail by many researchers from various points 
of view. For details, the reader may refer to the excellent reviews 
\cite{YE,SK}. The JCM is not only an important problem itself but also gives 
an excellent example of the so-called quantum open system problem\cite{D}, 
namely the interaction between a system and a reservoir.

So, we treated the JCM as a problem in non-equilibrium statistical mechanics 
and applied quantum mutual entropy\cite{O2} based on von Neumann entropy by 
finding the quantum mechanical channel\cite{O2} which expresses the state 
change of the atom on the JCM. This study was an attempt to obtain a new 
insight of the dynamical change of the state for the atom on the JCM by the 
quantum mutual entropy\cite{FO}.

On the other hand, this model has one of the most interesting features which 
is the entanglement developed between the atom and the field during the
interaction. There have been several approaches to analyze the time
evolution in this model, for instance, von Neumann entropy and atomic
inversion. Moreover, it is not suitable for the quantum mutual entropy which 
is mentioned above to measure the degree of entanglement. Because 
the quantum mutual entropy is defined for the quantum mechanical 
channel and when we obtain it, we take the partial trace over one system, 
then the entanglement tends to loss. Therefore, in this short paper, 
we will apply the DEM to analyze the entanglement of the time development of the JCM, since 
the DEM can measure the degree of entangled states directly without taking partial trace.

The resonant JCM Hamiltonian can be expressed by rotating-wave approximation
in the following form
\begin{eqnarray}
H =H_{A}+H_{F}+H_{I},  \nonumber \\
H_{A} ={\frac{1}{2}}\hbar \omega _{0}\sigma _{z},\,H_{F}=\hbar \omega
_{0}a^{*}a,  \nonumber \\
H_{I} =\hbar g(a\otimes \sigma ^{+}+a^{*}\otimes \sigma ^{-}),
\nonumber
\end{eqnarray}
where $g$ is a coupling constant, $\sigma ^{\pm }$ are the pseudo-spin
matrices of two-level atom, $\sigma _{z}$ is the $z$-component of the Pauli 
spin matrix, $a$ (resp. $a^{*}$) is the annihilation (resp. creation) 
operator of a photon.
In general, it is almost impossible to physically realize the pure states, so
we suppose the initial states of the atom is the mixed states which are the 
more realistic representation of the states.
We now suppose that the initial states of the atom are superposition states of the 
grounded states and the excited states:

$$\rho =\lambda _{0}E_{0}+\lambda _{1}E_{1}\in {\frak S}_{A}  $$
where $E_{0}=\left| 1\right\rangle \left\langle 1\right| $, $E_{1}=\left|
2\right\rangle \left\langle 2\right| $, $\lambda _{0}+\lambda _{1}=1$.
This means that we have a thermalized atom where there is no coherence 
between the levels.
Let the field initially be in squeezed states:
$$\omega =\left| \theta ; \xi \right\rangle \left\langle \theta ;\xi \right| 
\in {\frak S}
_{F},\,\left| \theta ;\xi \right\rangle =\exp \left( -\frac{1}{2}|\beta
|^{2}\right) \sum_{l}\frac{\beta ^{l}}{\sqrt{l!}}|l ;\xi\rangle , $$
where $\beta =\mu \theta +\nu \theta^* , \xi =r e^{i\theta}, \mu =\cosh r 
,\nu =\sinh r$ and $r$ is often called a squeezing parameter.
The continuous map ${\cal E}_{t}^{*}$, which is often called {\it lifting}\cite{AO} 
describing the time evolution between
the atom and the field for the JCM is defined by the unitary operator 
generated by $H$ such that
$${\cal E}_{t}^{*}:{\frak S}_{A}\longrightarrow {\frak S}_{A}\otimes {\frak 
S}
_{F},$$
\begin{equation}
{\cal E}_{t}^{*}\rho =U_{t}\left( \rho \otimes \omega \right) U_{t}^{*},
\label{ent_state}
\end{equation}
$$U_{t}\equiv \exp \left( -it\frac{H}{\hbar }\right).$$
This unitary operator $U_{t}$ is written as
\begin{equation}
U_{t}=\exp \left( {-itH/\hbar }\right) =\sum\limits_{n=0}^{\infty }{%
\sum\limits_{j=0}^{1}E_{n,j}\left| {\Phi _{j}^{\left( n\right) }}%
\right\rangle \left\langle {\Phi _{j}^{\left( n\right) }}\right| },
\label{unitary}
\end{equation}
where $E_{n,j}=\exp \left[ -it\left\{ \omega _{0}\left( n+\frac{1}{2}\right)
+\left( -1\right) ^{j}\Omega _{n}\right\} \right] $ are the eigenvalues with 
$
\Omega _{n}=g\sqrt{n+1},$ called Rabi frequency and ${\Phi _{j}^{\left(
n\right) }\ }$ are the eigenvectors associated to $E_{n,j}$.

The transition probability which the atom is initially prepared in the 
excited
state and stays at the excited state after the time $t$ is given by

$$ c\left( t \right)=\left| {\left\langle {n\otimes 2} \right|U_t\left| {%
n\otimes 2} \right\rangle } \right|^2  =\sum\limits_n{P\left(n\right) \cos 
^2 {\Omega _nt} }. $$

Also transition probability which the atom is initially prepared in the 
excited
state and is at the grounded state after the time $t$ is given by

$$ s\left( t \right)=\left| {\left\langle {n+1\otimes 1} \right|U_t\left| {%
n\otimes 2} \right\rangle } \right|^2  =\sum\limits_n{P\left(n\right) \sin 
^2 {\Omega _nt} }. $$
We note here that $P\left(n\right)$ is formulated by,
\begin{equation}
P\left(n\right) = \frac{1}{\vert \mu \vert n!}\left(\frac{\vert 
\nu\vert}{2\vert\mu\vert}\right)^n \left| H_n\left(\frac{\beta}{\sqrt{2 \mu 
\nu}}\right)\right|^2 \exp\left(-\vert \beta \vert^2+\frac{\nu}{2\mu}\beta^2 
+\frac{\nu^*}{2\mu}\beta^{*^2}\right) ,\label{subpoisson}
\end{equation}
where $\beta =\mu \theta +\nu \theta^*$ and $H_n\left(x\right)$ are Hermite 
polynomials.

\section{Derivation of DEM}

\quad \,
In this section, we derive the DEM for a two-level atom with squeezed state.
From (\ref{unitary}) and (\ref{subpoisson}), we obtain the lifting 
expression as follows;

\begin{eqnarray}
{\cal E}_{t}^{*}\rho =\sum_n{\left\{\lambda_0 P\left(n+1\right) \sin 
^2\Omega_n t +\lambda_1  P\left(n\right) \cos ^2\Omega_n t\right\}}\vert 2 
\rangle \langle 2\vert \otimes \vert n \rangle \langle n \vert
\nonumber
\\
+\frac{i}{2}\sum_n{\sin 2\Omega_n t\left(\lambda_1 P\left(n\right)-\lambda_0 
P\left(n+1\right)\right)}\vert 2 \rangle \langle 1\vert \otimes \vert n 
\rangle \langle n+1 \vert
\nonumber
\\
+\frac{i}{2}\sum_n{\sin 2\Omega_n t\left(\lambda_0 
P\left(n+1\right)-\lambda_1 P\left(n\right)\right)}\vert 1 \rangle \langle 
2\vert \otimes \vert n+1 \rangle \langle n \vert
\nonumber
\\
+\sum_n{\left\{\lambda_0 P\left(n\right) \sin ^2\Omega_n t +\lambda_1  
P\left(n+1\right) \cos ^2\Omega_n t\right\}}\vert 1 \rangle \langle 1\vert 
\otimes \vert n+1 \rangle \langle n+1 \vert
\end{eqnarray}

According to \cite{PK}, both atomic and field entropies are equal when the 
system is isolated each other, that is the final states are the pure states. 
However, since we have ${\cal E}_{t}^{*}\rho\neq \left({\cal E}_{t}^{*}\rho\right)^2$, the 
final states of the JCM are the entangled mixed states, so 
we should apply the DEM not the reduced von Neumann entropy.
Thus the DEM for a two-level atom with squeezed state is given by,

\begin{eqnarray}
I_{{\cal E}_{t}^{*}\rho}\left(\rho_t^A , \rho_t^F\right) = tr {\cal 
E}_{t}^{*}\rho \left(\log {\cal E}_{t}^{*}\rho - \log \rho_t^A \otimes \rho 
_t ^F\right)
\quad\qquad\qquad\qquad\quad\qquad\qquad\qquad\qquad\qquad
\nonumber \\
= -2\left( e_1\left(t\right) \log e_1\left(t\right) +e_4\left(t\right) \log 
e_4\left(t\right) \right)
+\kappa _+\left(t\right) \log \kappa _+\left(t\right)  + 
\kappa_-\left(t\right)  \log \kappa_-\left(t\right)
\end{eqnarray}
where,
\[
\kappa_{\pm} \left(t\right) = \frac{1}{2} \left\{ 
\left(e_1\left(t\right)+e_4\left(t\right)\right)\pm \sqrt{ 
\left(e_1\left(t\right)+e_4\left(t\right)\right)^2-4\left(e_1\left(t\right) 
e_4\left(t\right)- e_2\left(t\right) e_3\left(t\right) \right)}\right\},
\]
\begin{eqnarray*}
e_1\left(t\right)= \sum_n{\left\{\lambda_0 P\left(n+1\right) \sin ^2\Omega_n 
t +\lambda_1  P\left(n\right) \cos ^2\Omega_n t\right\}},
\\
e_2\left(t\right)=\frac{i}{2}\sum_n{\sin 2\Omega_n t\left(\lambda_1 
P\left(n\right)-\lambda_0 P\left(n+1\right)\right)} ,\\
e_3\left(t\right)=\frac{i}{2}\sum_n{\sin 2\Omega_n t\left(\lambda_0 
P\left(n+1\right)-\lambda_1 P\left(n\right)\right)} ,\\
e_4\left(t\right)= \sum_n{\left\{\lambda_0 P\left(n\right) \sin ^2\Omega_n t 
+\lambda_1  P\left(n+1\right) \cos ^2\Omega_n t\right\}}
\end{eqnarray*}

\section{Numerical computations}
\quad \,

On the basis of the analytical solution presented in the previous section, in this section
we examine the temporal evolution of the transition probability c(t), the probability distribution and the 
degree of entanglement for different values of the squeezed parameter r.

\begin{figure}[htbp]
\begin{center}
\includegraphics[width=5cm]{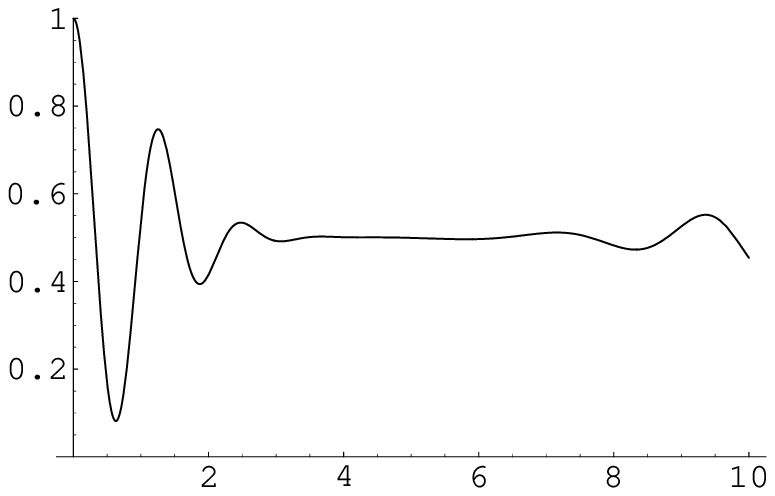}
\includegraphics[width=5cm]{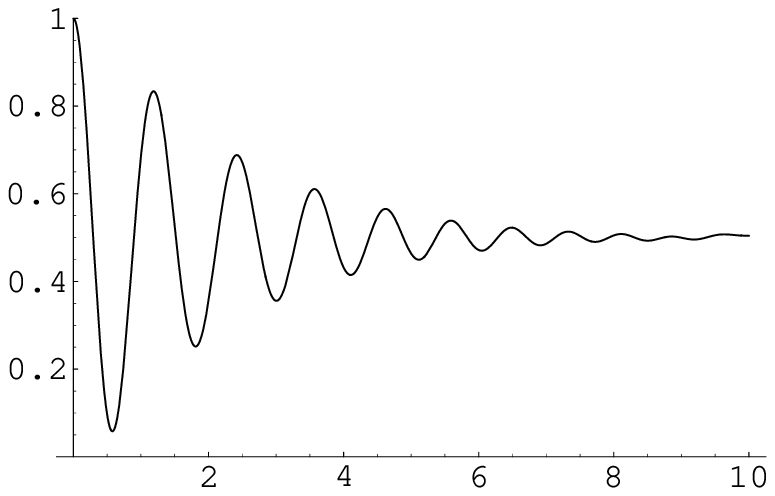}
\includegraphics[width=5cm]{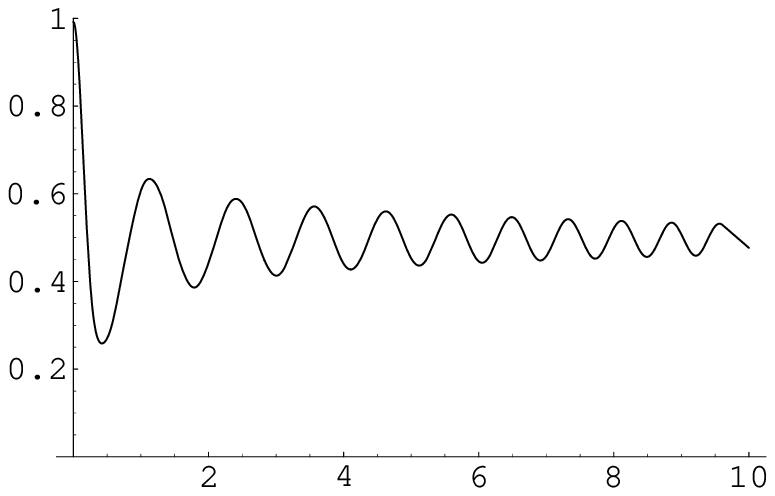}
\caption{$c\left(t\right)$ for time $t$ as $r=0$ (1a:left), $r=1$ (1b:center) and 
$r=2$ (1c:right).}
\end{center}
\end{figure}

We display the time evolution of the transition probability c(t) in
a two-level atom with squeezed states in figure 1,
since this measure has been often used to analyze the time development of 
the JCM. Figure 1a is obtained by setting the squeezing parameter $r=0$,
namely a coherent state which is a special case of the squeezed states and the 
change of the figure is well known as a nature of the coherent states 
JCM\cite{MS}.
To visualize the influence of the squeezed states
in the transition probability c(t)
we set different values of the squeezing parameter r (see Figure 1b, and 1c), while
all the other parameters are the same as in Figure 1a.
From these figures, it is remarkable that the frequency of the oscillations 
gradually gets to increase as a gain of the squeezing parameter $r$. However 
the size of the width of the amplitude in three figures is not monotone for 
a gain of the squeezing parameter $r$.

\begin{figure}[htbp]
\begin{center}
\includegraphics[width=7cm]{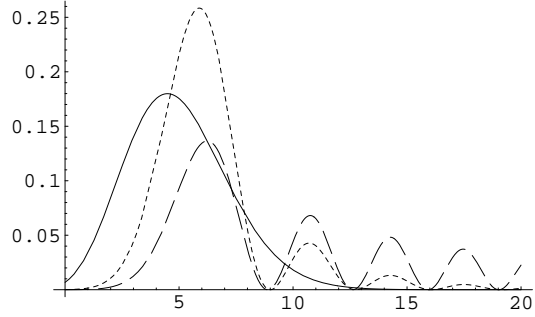}
\caption{The probability distribution $P(n)$ for $r=0$ (solid line) , $r=1$ 
(dotted line) and $r=2$ (dashed line).}
\end{center}
\end{figure}

These incomprehensible phenomena depend on the oscillation of the 
probability distribution of the squeezed state as increasing $r$\cite{WM} 
which the situation is also shown in the Figure 2 approximately described as 
a function of the photon number $n$ in the case of the mean photon number 
$\vert \theta \vert =\sqrt 5$.
It is known that the source of the oscillation is caused by the 
interference\cite{SW}. The effect of the squeezing for a two-level atom is 
examined by \cite{Mil}.

We would like to know how this property of the squeezing gives an influence 
of the entanglement of JCM.
Figure 3 shows the three-dimensional plot of the DEM as a function of
$\lambda_1$ and $r$ when the time $t$ is fixed in $t_r=2 \pi \theta /g$ 
which is often called the revival time. From this figure, we find that the 
DEM is not monotone for the squeezing parameter $r$ due to non-monotoneity 
of the amplitude of the transition probability for the squeezing parameter 
$r$.  Since the probability distribution functions in squeezed states oscillate
as the squeezing parameter $r$ (see figure 2), and it is known that this oscillation
is caused by the interference in phase space\cite{SW}.
From figure 3, we also find the degree of entanglement in $r=3$ is stronger 
than that in $r=0$, which means that by squeezing we can obtain the stronger 
entanglement than when we use the coherent states as the initial photon 
states, that is, it is possible to control the degree of the entanglement in 
the JCM by means of the squeezed states.

An interesting feature to observe in figure 3 is the symmetry around $\lambda_1=0.5$, which means that the DEM for the atom whether in ground or excited states are the same probability.
We also find by using the initial mixed states of the atom that the DEM always takes the maximum value in $\lambda_1 =0.5$ for any $r$ 
and the minimum value in $\lambda_1 =0$ or $1$ for any $r$. This result shows that we obtain the maximum degree of entanglement in the JCM when we use the most mixed states $\rho = 0.5 E_0 +0.5 E_1$ as the initial atomic state. This may be useful for the construction of quantum computer.

\begin{figure}[htbp]
\begin{center}
\includegraphics[width=7cm]{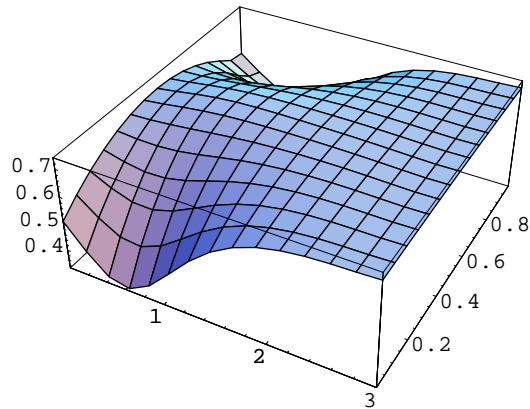}
\caption{The DEM for $\lambda_1 (0\leq \lambda_1 \leq 1)$ and $r (0\leq r\leq 3)$ when the time $t$ is fixed in 
$t_r=2 \pi \theta /g$}
\end{center}
\end{figure}

\section{Conclusion}
\quad \,

We have studied the influence of the squeezed states on
the degree of entanglement which is defined due to mutual entropy 
for a two-level atom.
This shows that the degree of entanglement is very sensitive to the
squeezing parameter.
For small values of  the squeezing parameter,
a decrease of the degree of entanglement is shown,
while for large values, an increase of the degree of entanglement
is obtained. This is manifested in the degree of entanglement
as it settles to a constant value for further increasing of the squeezing 
parameter.
This means that one can control the degree of entanglement by using the 
squeezing. We also found where the degree of entanglement in the JCM for any squeezing parameter $r$ 
takes the maximum value, applying the mixed states to the initial atomic states.
Moreover, we are interested in which the DEM has an upper bound for squeezing parameter $r$.
However, from only figure 3, we do not know whether the DEM has an upper bound or goes to infinity. 
This will remain as the forthcoming problem.

\section*{Acknowledgments}
\quad \,
The authors would like to thank the referee for his objective comments that improved the text in many points. We also would like to thank Prof. A.-S.F.Obada for his valuable discussion of 
this problem. Dr. S.Furuichi would like to thank the Grant-in-Aid for Encouragement of Young 
Scientists (No.11740078) from the Japanese Society for the Promotion of 
Science. 


\end{document}